\documentclass{article}
\usepackage{spconf}
\usepackage[noadjust]{cite}
\usepackage{bbm}
\usepackage{array}
\usepackage{dcolumn}
\usepackage{epsfig}
\usepackage[intlimits]{amsmath}
\usepackage{amsmath, amsfonts, yhmath, bm}
\usepackage{amssymb}
\usepackage{psfrag}
\usepackage{color,soul}
\usepackage[dvipsnames]{xcolor}
\usepackage[normalem]{ulem}
\usepackage{enumerate}
\usepackage{stackengine}
\usepackage{graphicx}
\usepackage{multirow}
\usepackage{etoolbox}
\usepackage{tcolorbox}
\usepackage{float}
\usepackage{dsfont}
\usepackage[ruled,vlined]{algorithm2e}
\usepackage{algpseudocode}
\usepackage{tikz}
\usepackage{mathtools}
\usepackage{arydshln}
\usetikzlibrary{shapes.geometric, arrows}

\usepackage{svg}
\usepackage{lipsum}  

\usepackage{hyperref}
\hypersetup{
    colorlinks,
    linkcolor={red!50!black},
    citecolor={blue!50!black},
    urlcolor={blue!80!black}
}

\usepackage{balance}
\usepackage{url}
\usepackage[inline]{enumitem} % For inline enumerations

\usepackage{vmr-symbols-vecbold}
\usepackage{standard-macros}
\PassOptionsToPackage{hyphens}{url}
\usepackage{hyperref}

\DeclareSymbolFontAlphabet{\amsmathbb}{AMSb}%

\newcommand{\lro}[1]{\lefto({#1}\right)}																% left right paranthesis operator
\newcommand{\lr}[1]{\left({#1}\right)}																% left right paranthesis operator
\newcommand{\lrb}[1]{\left \lbrace {#1} \right \rbrace}															% left right braces operator
\safemath{\dopplerspread}{B_D}																								% doppler spread
\safemath{\delayspread}{T_D}																									% delay spread
\safemath{\nc}{n\sub{c}}																										% coherence time
\safemath{\nf}{n\sub{f}}																										% feedback message length
\safemath{\efa}{p\sub{sc}}
\safemath{\efb}{p\sub{cs}}
\safemath{\ef}{\epsilon\sub{f}	}
\safemath{\nd}{n\sub{d}}																										% data symbols
\safemath{\ntx}{n\sub{t}} 																											% transmit antennas
\safemath{\nrx}{n\sub{r}}																											% receive antennas
\safemath{\ntxt}{\tilde{n\sub{t}}}																											% receive antennas
\safemath{\cb}{\ensuremath{L}} 																								% code blocks
\safemath{\cl}{\ensuremath{n}} 																								% codelength
\safemath{\txanto}{{\ensuremath{\tilde{m}_t}}} 																		% transmit antennas when some is turned off
\safemath{\cs}{M} 																														% code size
\safemath{\idPustm}{\ensuremath{S_{k}}}
\safemath{\error}{\ensuremath{\epsilon}} 																				%Error target
\safemath{\eexp}{\ensuremath{\mathcal{E}}} 																			%Error exponent
\safemath{\nsubc}{n\sub{s}}			 																						% number of subcarriers
\safemath{\nofdm}{n\sub{o}} 																									% number of OFDM symbols
\safemath{\bc}{\ensuremath{B_c}} 																							% Coherence bandwidth
\safemath{\ts}{\ensuremath{T_s}} 																							% Symbol time
\safemath{\nrb}{\ensuremath{n_{rb}}} 																						% Symbol time
\safemath{\rul}{\ensuremath{\rho\sub{ul}}}
\safemath{\rdl}{\ensuremath{\rho\sub{dl}}}

\safemath{\nres}{\ell}
\safemath{\nr}{n\sub{r}}
\safemath{\maxk}{M^*\lr{\nres, \nsubc, \nofdm, \epsilon, \rho}}
\safemath{\Rmax}{R^*}%\lr{\nres, \nsubc, \nofdm,M, \epsilon, \rho}}
\safemath{\Emin}{E\sub{b}^*/N_0}%\lr{\nres, \nsubc, \nofdm,M, \epsilon, \rho}}
\safemath{\Eminf}{\frac{E\sub{b}^*}{N_0}}
\safemath{\np}{\ensuremath{n\sub{p}}}
\safemath{\ndf}{\ensuremath{\bar{n}\sub{d}}}
\safemath{\npf}{\ensuremath{\bar{n}\sub{p}}}
\safemath{\code}{\ensuremath{\mathcal{C}}}
\safemath{\err}{\ensuremath{\epsilon}}
\safemath{\rp}{\ensuremath{\rho\sub{p}}}
\safemath{\rd}{\ensuremath{\rho\sub{d}}}
\safemath{\cohtime}{\ensuremath{T\sub{c}}}
\safemath{\cohbw}{\ensuremath{B\sub{c}}}
\safemath{\nmax}{\ensuremath{\ell\sub{m}}}
\safemath{\ntot}{\ensuremath{n\sub{tot}}}
\safemath{\nul}{\ensuremath{n\sub{ul}}}
\safemath{\ndl}{\ensuremath{n\sub{dl}}}

\safemath{\yp}{\ensuremath{\randvecy_{\nu}^{(\text{p})}}}
\safemath{\yd}{\ensuremath{\randvecy_{\nu}^{(\text{d})}}}
\safemath{\ypd}{\ensuremath{\vecy_{\nu}^{(\text{p})}}}
\safemath{\ydd}{\ensuremath{\vecy_{\nu}^{(\text{d})}}}

\safemath{\ypf}{\ensuremath{\bar{\randvecy}_{\nu}^{(\text{p})}}}
\safemath{\ydf}{\ensuremath{\bar{\randvecy}_{\nu}^{(\text{d})}}}
\safemath{\ypdf}{\ensuremath{\bar{\vecy}_{\nu}^{(\text{p})}}}
\safemath{\yddf}{\ensuremath{\bar{\vecy}_{\nu}^{(\text{d})}}}

\safemath{\xp}{\ensuremath{\vecx^{(\text{p})}}}
\safemath{\xd}{\ensuremath{\randvecx_{\nu}^{(\text{d})}}}
\safemath{\xdd}{\ensuremath{\vecx_{\nu}^{(\text{d})}}}

\safemath{\xpf}{\ensuremath{\bar{\vecx}^{(\text{p})}}}
\safemath{\xdf}{\ensuremath{\bar{\randvecx}_{\nu}^{(\text{d})}}}
\safemath{\xddf}{\ensuremath{\bar{\vecx}_{\nu}^{(\text{d})}}}

\safemath{\xdb}{\ensuremath{\overline{\randvecx}^{(\text{d})}}}
\safemath{\Pxd}{\ensuremath{P_{\randvecx^{(\text{d})}}}}

\safemath{\xpbar}{\ensuremath{\overline{\matX}^{(\text{p})}}}
\safemath{\xdbar}{\ensuremath{\overline{\randmatX}^{(\text{d})}}}

\safemath{\xdv}{\ensuremath{\randvecx^{(\text{d})}}}
\safemath{\xdbarv}{\ensuremath{\overline{\randvecx}^{(\text{d})}}}
\safemath{\ydv}{\ensuremath{\randvecy^{(\text{d})}}}

\safemath{\xdr}{\ensuremath{\matX^{(\text{d})}}}

\safemath{\ttx}{\ensuremath{\tau\sub{tx}}}
\safemath{\trx}{\ensuremath{\tau\sub{rx}}}
\safemath{\ack}{\ensuremath{\mathrm{s}}}
\safemath{\nack}{\ensuremath{\mathrm{c}}}

\newcommand{\prob}[1]{\ensuremath{\mathbb{P}\lro{#1}}}

\safemath{\mI}{\ensuremath{i\lro{\randvecy ; \randvecx}}} 				% i(Y;X)
\safemath{\randveca}{\bm{A}}
\safemath{\randvecb}{\bm{B}}
\safemath{\randvecc}{\bm{C}}
\safemath{\randvecd}{\bm{D}}
\safemath{\randvece}{\bm{E}}
\safemath{\randvecf}{\bm{F}}
\safemath{\randvecg}{\bm{G}}
\safemath{\randvech}{\bm{H}}
\safemath{\randveci}{\bm{I}}
\safemath{\randvecj}{\bm{J}}
\safemath{\randveck}{\bm{K}}
\safemath{\randvecl}{\bm{L}}
\safemath{\randvecm}{\bm{M}}
\safemath{\randvecn}{\bm{N}}
\safemath{\randveco}{\bm{O}}
\safemath{\randvecp}{\bm{P}}
\safemath{\randvecq}{\bm{Q}}
\safemath{\randvecr}{\bm{R}}
\safemath{\randvecs}{\bm{S}}
\safemath{\randvect}{\bm{T}}
\safemath{\randvecu}{\bm{U}}
\safemath{\randvecv}{\bm{V}}
\safemath{\randvecw}{\bm{W}}
\safemath{\randvecx}{\bm{X}}
\safemath{\randvecy}{\bm{Y}}
\safemath{\randvecz}{\bm{Z}}
\safemath{\randvecphi}{\bm{\Phi}}

\safemath{\randmatA}{\amsmathbb{A}}
\safemath{\randmatB}{\amsmathbb{B}}
\safemath{\randmatC}{\amsmathbb{C}}
\safemath{\randmatD}{\amsmathbb{D}}
\safemath{\randmatE}{\amsmathbb{E}}
\safemath{\randmatF}{\amsmathbb{F}}
\safemath{\randmatG}{\amsmathbb{G}}
\safemath{\randmatH}{\amsmathbb{H}}
\safemath{\randmatI}{\amsmathbb{I}}
\safemath{\randmatJ}{\amsmathbb{J}}
\safemath{\randmatK}{\amsmathbb{K}}
\safemath{\randmatL}{\amsmathbb{L}}
\safemath{\randmatM}{\amsmathbb{M}}
\safemath{\randmatN}{\amsmathbb{N}}
\safemath{\randmatO}{\amsmathbb{O}}
\safemath{\randmatP}{\amsmathbb{P}}
\safemath{\randmatQ}{\amsmathbb{Q}}
\safemath{\randmatR}{\amsmathbb{R}}
\safemath{\randmatS}{\amsmathbb{S}}
\safemath{\randmatT}{\amsmathbb{T}}
\safemath{\randmatU}{\amsmathbb{U}}
\safemath{\randmatV}{\amsmathbb{V}}
\safemath{\randmatW}{\amsmathbb{W}}
\safemath{\randmatX}{\amsmathbb{X}}
\safemath{\randmatY}{\amsmathbb{Y}}
\safemath{\randmatZ}{\amsmathbb{Z}}
\safemath{\randmatSigma}{\mathbb{\Sigma}}
\safemath{\randmatPhi}{\mathbb{\Phi}}
\safemath{\randmatLambda}{\mathbb{\Lambda}}

\safemath{\matSigma}{\bm{\Sigma}}
\safemath{\matPhi}{\bm{\Phi}}
\safemath{\matLambda}{\bm{\Lambda}}

\usepackage{glossaries}
\glsdisablehyper
\newglossaryentry{mie}
{
  name={MIE},
  description={maximum index-based estimator},
  first={\glsentrydesc{mie} (\glsentrytext{mie})}
}
\usepackage{pgfplots}
\pgfplotsset{compat=1.14}
\usepgfplotslibrary{groupplots}
\usetikzlibrary{pgfplots.groupplots}

\newtheorem{theorem}{Theorem}
\newtheorem{lemma}{Lemma}
\newtheorem{corollary}{Corollary}

\newtheorem{proposition}{Proposition}

\newcommand{\mle}{\text{\tiny MLE}}
\newcommand{\mie}{\text{\tiny MIE}}
\newcommand{\ratedist}{\text{\tiny RD}}
\newcommand{\onebit}{\text{\tiny 1-bit}}

\newcommand {\snr} {\mathtt{SNR}}

\makeatletter
\newcommand\footnoteref[1]{\protected@xdef\@thefnmark{\ref{#1}}\@footnotemark}
\makeatother
\newcommand\barbelow[1]{\stackunder[1.2pt]{$#1$}{\rule{.8ex}{.075ex}}}
\usepackage[lining]{ebgaramond}

\makeatletter
\newsavebox\myboxA
\newsavebox\myboxB
\newlength\mylenA

\newcommand*\mybar[2][0.75]{%
	\sbox{\myboxA}{$\m@th#2$}%
	\setbox\myboxB\null% Phantom box
	\ht\myboxB=\ht\myboxA%
	\dp\myboxB=\dp\myboxA%
	\wd\myboxB=#1\wd\myboxA% Scale phantom
	\sbox\myboxB{$\m@th\overline{\copy\myboxB}$}%  Overlined phantom
	\setlength\mylenA{\the\wd\myboxA}%   calc width diff
	\addtolength\mylenA{-\the\wd\myboxB}%
	\ifdim\wd\myboxB<\wd\myboxA%
	\rlap{\hskip 0.5\mylenA\usebox\myboxB}{\usebox\myboxA}%
	\else
	\hskip -0.5\mylenA\rlap{\usebox\myboxA}{\hskip 0.5\mylenA\usebox\myboxB}%
	\fi}
\makeatother
\makeatletter
\newcommand\customsize{\@setfontsize\customsize{11}{13.6}}
\makeatother
\title{A Joint Data Compression and Time-Delay Estimation Method For Distributed Systems via Extremum Encoding}

\name{Amir Weiss$^{\star}$,  Yuval Kochman$^{\dagger}$ and Gregory W. Wornell$^{\ddagger}$}

\address{\customsize
\begin{tabular}{ccc}
$^{\star}$Faculty of Engineering & $^\dagger$School of Computer Science and Engineering & $^{\ddagger}$Research Laboratory of Electronics\\
Bar-Ilan University & The Hebrew University of Jerusalem & Massachusetts Institute of Technology\\
amir.weiss@biu.co.il & yuvalko@cs.huji.ac.il  & gww@mit.edu
\end{tabular}
}

\begin{document}
\ninept
\maketitle

\begin{abstract}
Motivated by the proliferation of mobile devices, we consider a basic form of the ubiquitous problem of time-delay estimation (TDE), but with communication constraints between two non co-located sensors. In this setting, when joint processing of the received signals is not possible, a compression technique that is tailored to TDE is desirable. For our basic TDE formulation, we develop such a joint compression-estimation strategy based on the notion of what we term ``extremum encoding'', whereby we send the index of the maximum of a finite-length time-series from one sensor to another. Subsequent joint processing of the encoded message with locally observed data gives rise to our proposed time-delay ``maximum-index''-based estimator. We derive an exponentially tight upper bound on its error probability, establishing its consistency with respect to the number of transmitted bits. We further validate our analysis via simulations, and comment on potential extensions and generalizations of the basic methodology.

\end{abstract}
\vspace{-0.175cm}
\section{Introduction}\label{sec:intro}
\vspace{-0.15cm}
Time-delay estimation (TDE) is a fundamental problem that is found at the core of numerous applications in various scientific fields and physical domains (e.g., acoustic, optics, radio frequency). Examples, among others, include localization, tracking, communication, sensor calibration, medical imaging and more \cite{viola2003comparison,li2006position,musicki2009mobile,weiss2022semi}. In this respect, it is perhaps one of the most important problems in signal processing, and as such, it has been extensively studied in past decades. For a collection of important results, which nevertheless does not serve as an exhaustive survey, see \cite{ziv1969some,quazi1981overview,ianniello1982time,weiss1983fundamental,weinstein1984fundamental,azaria1984time,fertner1986comparison,carter1987coherence,jacovitti1993discrete,brandstein1997robust,bjorklund2003survey,benesty2004time,chen2006time}.

Driven by recent technological developments  \cite{da2014internet}, in a growing number of settings bandwidth constraints necessitate the use of data compression when carrying out TDE, but such considerations have received comparatively less attention. In classical settings, it is typically assumed that the central computing unit has access to both of the received signals. This is a reasonable assumption when the sensors are co-located or when there are no constraints on the relevant communication links. In contrast, for some of the modern emerging applications, this is no longer the case. Consider, for example, the problem of passive acoustic indoor localization \cite{liu2020indoor}, building on power- and communication-limited ``smart" devices. We envision that in this type of applications, these devices would opportunistically be used as \emph{ad-hoc} sensors that could---with limited resources---measure an acoustic signal and convey a corresponding message for the purpose of TDE (e.g., as a proxy for range estimation, or more generally, as a building block in distributed localization systems).

In such scenarios, it is not only desirable, but already necessary to reduce as much as possible the resources requirements on the spatially distributed devices that serve as the (low-cost) receivers. Note that, clearly, the sensors in this case are not co-located. Moreover, the assumption of an essentially unlimited communication link is weak in some cases and unrealistic in others. A similar set of constraints can be presented by sensor networks \cite{akyildiz2002survey,chen2002source}, which by design consist of a large number of small, low-power, untethered devices that measure a common signal for the collective purpose of some inference task (e.g., \cite{wang2003preprocessing}).

This motivation, with the proliferation of wireless devices, has led to several works, focused on compression techniques that are specifically designed for TDE. A few representative examples are \cite{vasudevan2003application,fowler2005fisher,chen2010data,fuyong2012data,vargas2018compressed}. A common theme in these previous works is that the corresponding proposed methods are eventually trying to \emph{best compress the received signal} to be sent to the central computing unit (often, a different receiver). However, this conceptual limitation is not a must. Instead, one may design a \emph{joint} compression-estimation scheme in which the compression produces some message that is \emph{best matched to the final (and specific) estimation task}, and is perhaps only an extremely coarsely compressed version of the received signal itself. Our idea is based exactly on this notion, and consequently opens the search space to a larger set of potential solutions.

In this work, we consider a basic, slightly simplified formulation of the TDE problem, but with communication constraints for two non co-located sensors, where joint processing of the two signals is not possible. We focus on theoretical aspects thereof, and establish key preliminary---conceptual and analytical---results that pave the way to additional, more practical extensions, which can be further developed based on our results.

\vspace{-0.3cm}
\subsection{Contributions}\label{subsec:contributions}
% Motivated by the above, and inspired by information theoretic ideas presented in recent work by Hadar and Shayevitz \cite{hadar2019distributed} and Kochman and Wang \cite{kochman2021communication}, we propose a new paradigm for distributed TDE with communication constraints. While the present work is focused on a specific signal model, the conceptual insight stemming from it is of both theoretical and practical value for the design for compression methods for TDE in other different settings that are similar in nature.
\vspace{-0.1cm}
% Specifically, our main contributions are the following:
Motivated by the above, and inspired by information theoretic ideas presented in recent works \cite{hadar2019distributed}, \cite{kochman2021communication} on distributed inference of correlation between sequences without delay, our main contributions are as follows:
\begin{itemize}[itemsep=0.15pt]%[noitemsep,nolistsep]
    \item \emph{A new method for joint compression and TDE for distributed systems with communication constraints}: Our proposed method is based upon sending only the index of the maximal observed value in some prescribed range, and is therefore computationally simple and can be interpreted intuitively. It is demonstrated via a simulation experiment to be favorable relative to two benchmarks in terms of the inherent trade-off between the number of bits sent and estimation fidelity. %\textbf{This last statement should be phrased gently, as we discussed last time}\awedit{I agree, but we are just saying benchmarks, not ``state-of-the-art" or someting like that}
    % \item \emph{Asymptotic Performance Analysis}: We derive the asymptotic rate of decay of the error probability of our estimator, and in particular show that it is consistent, i.e., has a vanishing error probability with an increasingly long observation time.
    \item \emph{Performance Analysis}: We derive an explicit upper bound on the error probability of our scheme, and show that it decays exponentially as a function of the number of bits sent. Thus, in particular, it is consistent in the communication sense. We further show that this exponential behaviour is tight for our scheme. 
    
\end{itemize}

\vspace{-0.38cm}
\section{Problem Formulation}\label{sec:problemformulation}
\vspace{-0.2cm}
We start by formulating a simplified version of the TDE problem. In particular, consider the observed discrete-time signals at two distant sensors 
\begin{equation}\label{eq:initialmodel}
\begin{aligned}
    \rndr_1[n] &= \rndx[n] + \rndz_1[n], \;\; &\text{(sensor 1)} \\
    \rndr_2[n] &= \rndx[n-\rndd] + \rndz_2[n], \;\; &\text{(sensor 2)}
\end{aligned}
\end{equation}
where
\begin{itemize}
    \item $\rndx[n]\overset{\text{iid}}{\sim}\normal(0,1)$ is the common signal that is observed by both sensors with a relative time-delay $\rndd\in\setD$, where $\setD\triangleq\{-d_m,\ldots,d_m\}$ is the ``uncertainty interval" (or the ``delay spread") and $d_m\in\naturals$ is the maximum (absolute) delay; and
    \item $\rndz_1[n]\overset{\text{iid}}{\sim} \normal(0,\sigma_1^2), \rndz_2[n]\overset{\text{iid}}{\sim} \normal(0,\sigma_2^2)$ are statistically independent white Gaussian noise processes with unknown deterministic variances $\sigma_1^2, \sigma_2^2$, which are also statistically independent of $\rndx[n]$.
    % \item $\rndz_1[n],\rndz_2[n]\overset{\text{iid}}{\sim} \normal(0,1)$ are statistically independent additive white Gaussian noise processes, that are also statistically independent of $\rndx[n]$, and $\sigma_1,\sigma_1\in\positivereals$ are unknown constants.
\end{itemize}

\begin{figure}
    \centering
    \includegraphics[width=0.9\columnwidth]{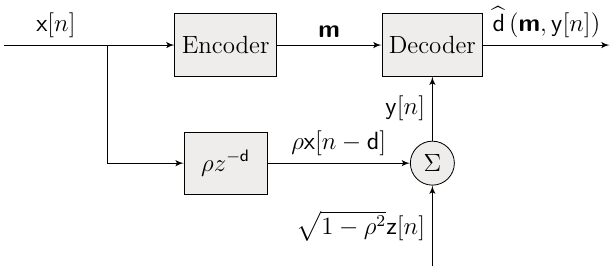}\vspace{-0.2cm}
    \caption{Abstraction of the distributed time-delay estimation problem considered in this work. The encoder observes $\rndx[n]$ and generates a message of length $k$ bits $\rvecm\in\{0,1\}^{k\times 1}$. The decoder observes $\rndy[n]$ and receives $\rvecm$, from which it constructs $\widehat{\rndd}(\rvecm,\rndy[n])$, an estimator of $\rndd$.}
    \label{fig:blockdiagram}\vspace{-0.4cm}
\end{figure}
For ease of notation, we begin with the following proposition that allows us to continue the analysis with a simplification of the model.\vspace{-0.2cm}
\begin{proposition}\label{proposition1}
Model \eqref{eq:initialmodel} is statistically equivalent to the model
\begin{equation}\label{eq:model}
\begin{aligned}
    \rndx[n]&, &\text{\emph{(sensor 1, ``encoder")}} \\
    \rndy[n] &= \rho\rndx[n-\rndd] + \bar{\rho}\rndz[n], &\text{\emph{(sensor 2, ``decoder")}}
\end{aligned}
\end{equation}
depicted in Fig.~\ref{fig:blockdiagram}, where $\rndx[n]\overset{\text{\emph{iid}}}{\sim} \normal(0,1)$ and $\rndz[n]\overset{\text{\emph{iid}}}{\sim} \normal(0,1)$ are statistically independent, $\rho\in(0,1]$ is the (Pearson) correlation coefficient between $\rndx[n]$ and $\rndy[n+\rndd]$ that is related to the signal-to-noise ratios (SNRs) of \eqref{eq:initialmodel}, i.e., to $1/\sigma_1^2$ and $1/\sigma_2^2$, and $\bar{\rho}\triangleq\sqrt{1-\rho^2}$.
\end{proposition}
The (simple) proof is omitted due to space considerations, but it is easy to see that, up to power (/variance) normalization of the observed signals, the processes in both pairs $\rndr_1[n],\rndr_2[n]$ and $\rndx[n],\rndy[n]$ are each marginally white Gaussian, and are also jointly Gaussian with an identical (up to scaling) cross-correlation.
Therefore, we will henceforth work with the observation model \eqref{eq:model}, and accordingly, we shall refer to the quantity $\snr\triangleq\frac{\rho^2}{\bar{\rho}}=\frac{\rho^2}{1-\rho^2}$ (with which $\rho^2=\frac{1}{1+\snr^{-1}}$) as the SNR.\footnote{Observe that $\snr\xrightarrow[]{\rho\to1}\infty$ and $\snr\xrightarrow[]{\rho\to0}0$, as desired.}% For the exact relation between $\snr$ and $1/\sigma_1^2,1/\sigma_2^2$ (SNRs in model \eqref{eq:initialmodel}).

In \eqref{eq:model}, one sensor (the ``encoder") observes\footnote{With a slight abuse of notation, we write that ``one observes $\rndx[n]$" when we mean that one observes the entire process $\{\rndx[n]\}_{n\in\integers}$ or a snippet of it.} $\rndx[n]$ and needs to produce a message $\rvecm\in\{0,1\}^{k\times 1}$ of length $k\in\naturals$ bits to be sent to the other sensor (the ``decoder"). The latter observes $\rndy[n]$, a noisy version of (the $\rho$-scaled) $\rndx[n]$, delayed by $\rndd$ samples. We consider a Bayesian setting in which $\rndd\sim\mathcal{U}(\setD)$ (i.e., uniformly distributed), and the goal of the sensor 2, which is assumed to be located at the central computing unit, is to estimate $\rndd$ based on the observed signal $\rndy[n]$ and the received message $\rvecm$, so as to minimize the risk $\Exop\left[\ell(\widehat{\rndd},\rndd)\right]$ for a loss function $\ell:\setD\times\setD\to\positivereals$, where the expectation is with respect to all sources of randomness, i.e., $\rndx[n],\rndz[n]$ and $\rndd$. In this work, we focus on the error loss $\ell(a,b)=\mathbbm{1}_{a\neq b}$ that yields the error probability risk. We are interested in the trade-off between the number of bits $k$ and the error probability.
% \vspace{0.2cm}
\section{A Joint Compression-Estimation Scheme}\label{sec:compressestimate}
% \vspace{-0.2cm}
Our proposed strategy, which is inspired by \cite{hadar2019distributed} and \cite{kochman2021communication}, is as follows. 

\textbf{Extremum Encoding:} The encoder observes a sequence of length $N=2^{k}$,\footnote{This assumption is merely for notational convenience, and can of course be relaxed, in the sense that $N$ can be any natural number.} namely $\mathcal{X}_N\triangleq\{\rndx[n]\}_{n=0}^{N-1}$, and sends as the message $\rvecm$ (of length $k$ bits) the \emph{index} of the maximum sample among all $\mathcal{X}_N$,
\begin{equation}\label{definitionofJ}
    \rndj\triangleq \arg \max_{0\leq n\leq N-1} \; \rndx[n],% \; \Longrightarrow \; \rvecm= {\rm{dec\_to\_bin}}\left(\rndj\right),
\end{equation}
where $\rvecm\in\{0,1\}^{k\times 1}$ is the binary representation of $\rndj$.
% where ${\rm{dec\_to\_bin}}\left(\cdot\right)$ converts an index (natural number) into a bit string corresponding to its binary representation.
% Clearly, the message $\rvecm$ can by conveyed with $nk$, as required.

\textbf{Maximum-Index-Based Estimation:} The decoder, which in particular observes $\mathcal{Y}_{N}^{\setD}\triangleq\{\rndy[n]\}_{n=-d_m}^{N-1+d_m}$, upon receiving the message $\rvecm$ (equivalently, the index $\rndj$), constructs% the following estimator,
\begin{equation}\label{eq:maxestsimple}
{\widehat{\rndd}}_{\mie} \triangleq \arg\max_{\ell\in\setD} \; \rndy[\rndj+\ell],
\end{equation}
which we call the ``maximum-index"-based estimator (MIE). Put simply, the message from the encoder simply dictates to the decoder the center of its search (discrete-)time interval, whose size is the delay spread, i.e., $|\setD|=2d_m+1$. The estimated time-delay is then chosen to be the shift (in the opposite direction, c.f.\ \eqref{eq:model}) relative to that center, for which the observed signal at the decoder is maximized.

\subsection{Interpretation of the MIE}\label{subsec:interpretation}
The underlying logic of \eqref{eq:maxestsimple} is in fact quite intuitive. To reveal it, let us first consider the maximum \textit{a posteriori} estimator of $\rndd$ from $\rndx[n]$ and $\rndy[n]$, which in our case (since $\rndd\sim\setU(\setD)$) coincides with the maximum likelihood estimator (MLE), and can be easily shown to be given by
% It is easy to show that the maximum \textit{a posteriori} estimator of $\rndd$, which coincides with the maximum likelihood estimator (MLE) in our case since $\rndd\sim\setU(\setD)$, is given by
\begin{equation}\label{eq:mledefinition}
    \widehat{\rndd}_{\mle} = \arg\max_{\ell\in\setD} \; \frac{1}{N}\sum_{n=0}^{N-1}\rndx[n]\rndy[n+\ell] \triangleq \arg\max_{\ell\in\setD} \; \widehat{\rho}_{\mle}(\ell).
\end{equation}
Although by different means, the MIE \eqref{eq:maxestsimple} is doing exactly what the MLE \eqref{eq:mledefinition} is doing without communication constraints, which is simply trying to identify the time-lag at which the cross-correlation between $\rndx[n]$ and $\rndy[n]$ is maximized. To see this clearly, we recall the following result.

\noindent\textbf{Theorem (Hadar and Shayevitz, \cite{hadar2019distributed})} \textit{Consider model \eqref{eq:model} with $\rndd\equiv0$, namely, $\{\rndx[n]\}$ and $\{\rndy[n]\}$ are zero-mean unit-variance white (i.e., uncorrelated) Gaussian processes with a correlation coefficient $\rho$. Let 
\begin{equation}
\widehat{\rho}_{\emph{\mie}} \triangleq \frac{\rndy[\rndj]}{\Exop[\rndx[\rndj]]},    
\end{equation}
where $\rndj$ is defined in \eqref{definitionofJ}. Then, $\widehat{\rho}_{\emph{\mie}}$ is an unbiased estimator of $\rho$ with
\begin{equation}\label{eq:varofcorrelationmaxestimator}
\Varop\left({\widehat{\rho}}_{\emph{\mie}}\right) = \frac{1}{k}\left(\frac{1-\rho^2}{2\log(2)}+o(1)\right) = \frac{1-\rho^2}{2\log(N)}+o(1).
\end{equation}
Moreover, $\widehat{\rho}_{\emph{\mie}}$ is asymptotically efficient\footnote{An efficient estimator is an unbiased estimator that attains the Cram\'er-Rao lower bound \cite{van2004detection}.} given $(\rndx[\rndj],\rndy[\rndj])$.}

A natural extension (/application) of the above is to define
\begin{equation}\label{eq:correlationmaxestimator}
\widehat{\rho}_{\mie}(\ell) \triangleq \frac{\rndy[\rndj+\ell]}{\Exop\left[\rndx[\rndj]\right]}, \quad \forall \ell\in\integers,
\end{equation}
which is of course an unbiased, asymptotically efficient estimator of $\rho$ when $\ell=\rndd$, and of $0$ when $\ell\neq\rndd$. This is simply because, for any time shift $\ell$, we end up with exactly the same formulation considered in \cite{hadar2019distributed}, but for a different correlation coefficient ($\rho$ or zero, as explained above).

With \eqref{eq:correlationmaxestimator}, we can revisit \eqref{eq:maxestsimple}, and using the fact that $\Exop[\rndx[\rndj]]$ is constant with respect to the optimization index $\ell$, we may write
{\setlength{\belowdisplayskip}{5pt} \setlength{\belowdisplayshortskip}{5pt}
\setlength{\abovedisplayskip}{5pt} \setlength{\abovedisplayshortskip}{5pt}
\begin{equation}\label{eq:maxestsimpleascorr}
{\widehat{\rndd}}_{\mie} = \arg\max_{\ell\in\setD} \; \frac{\rndy[\rndj+\ell]}{\Exop\left[\rndx[\rndj]\right]}
    =\arg\max_{\ell\in\setD} \; \widehat{\rho}_{\mie}(\ell).
\end{equation}}
Indeed, it is now evident that the MIE \eqref{eq:maxestsimpleascorr} (for limited communication) and the MLE \eqref{eq:mledefinition} (for unlimited communication) are similar in nature---both are choosing the hypothesized time-lag at which their respective empirical cross-correlations are maximized as the estimated time-delay.
\vspace{-0.25cm}
\subsection{Computational Complexity}\label{subsec:somplexity}
\vspace{-0.15cm}Beyond its asymptotic performance (discussed in Section \ref{sec:asymptoticperformance} below), another appealing property of the MIE \eqref{eq:maxestsimple} is its complexity, in particular relative to standard cross-correlation-based estimators, such as the MLE \eqref{eq:mledefinition}. For such standard cross-correlators, computing the empirical cross-correlation at $|\setD|=2d_m+1$ time-lags based on $N$ samples amounts to $\setO(Nd_m)$ operations. In contrast, the MIE simply requires two searches for the maximum of two arrays of sizes $N$ (encoder) and $|\setD|$ (decoder), hence its complexity is $\setO(N+d_m)$. It is therefore evident that our proposed method is not only more efficient in terms of communication, but is also attractive in terms of the computational resources it requires.
\vspace{-0.2cm}
\section{Performance Analysis of the MIE}\label{sec:asymptoticperformance}
\vspace{-0.15cm}
While the intuition of \eqref{eq:maxestsimple} provided in Section \ref{subsec:interpretation} is reassuring, it is imperative to also provide the accompanying analytical performance guarantees. Fortunately, due to the asymptotic concentration of $\rndx[\rndj]$---the maximum of a finite-length realization of a white Gaussian process---around its mean, the intuitive interpretation of \eqref{eq:maxestsimple} presented above can also be rigorously justified. In particular, we have the following result regarding the error probability of the MIE, whose proof appears in Section \ref{subsec:proofs}.
\begin{theorem}[Error probability upper and lower bounds]\label{theorem1}
Consider the error loss $\ell(a,b)=\mathbbm{1}_{a\neq b}$, giving the error probability risk $\epsilon\triangleq\prob{\widehat{\rndd}_{\emph{\mie}}\neq \rndd}$. Then, for a sufficiently large $k$ and any $\rho\in(0,1]$,\footnote{The case $\rho=0$ is less interesting, and trivial to analyze, since both sensors observe (purely) statistically independent white Gaussian noise.}% we have
\begin{align}
    \barbelow{\epsilon}(k,\rho)\left(1+o(1)\right) \leq \epsilon \leq \bar{\epsilon}(k,\rho,d_m)\left(1+o(1)\right)\label{eq:upperandlowerbound},
\end{align}
where
\begin{align}
    \bar{\epsilon}(k,\rho,d_m) &\triangleq 2d_m\cdot e^{-k\frac{\log(2)\cdot\rho^2}{2-\rho^2}}, \label{eq:upperbound}\\
    \barbelow{\epsilon}(k,\rho) &\triangleq \sqrt{\frac{2-\rho^2}{4\pi\rho^2\log(2)k}}\cdot e^{-k\frac{\log(2)\cdot\rho^2}{2-\rho^2}},\label{eq:lowerbound}
\end{align}
% \begin{align}
%     \epsilon&\leq \frac{2d_m}{\sqrt{2-\rho^2}}\cdot\frac{1}{N^{\frac{\rho^2}{2-\rho^2}}}\left(1+o(1)\right)\triangleq \bar{\epsilon}(N,\rho,d_m)\label{eq:upperboundmaintheoremdiscretetime},
% \end{align}
and where the $o(1)$ term goes to zero as $k\to\infty$.
\end{theorem}
\vspace{-0.15cm}
An immediate corollary of Theorem \ref{theorem1} is the following.\vspace{-0.15cm}
\begin{corollary}[``Communication Consistency"]\label{corollary1}
For the same setting of Theorem \ref{theorem1}, the MIE \eqref{eq:maxestsimple} is consistent in the communication sense, namely,
\begin{equation}\label{eq:consistencyofmie}
    \lim_{k\to\infty}\prob{\widehat{\rndd}_{\emph{\mie}}\neq \rndd}=0.
\end{equation}
Moreover, $d_m$ need \emph{not} be fixed, and it is only required that $d_m=o\left(\exp\left\{k\frac{\log(2)\cdot\rho^2}{2-\rho^2}\right\}\right)$. In other words, the delay spread can grow with the observation time, as long as it grows ``sufficiently slow" with $k$. Additionally, we note in passing that $\widehat{\rndd}_{\emph{\mie}}$ is also consistent with respect to the number of samples that need to be used, $N$, i.e., $\prob{\widehat{\rndd}_{\emph{\mie}}\neq \rndd}\xrightarrow[]{N\to\infty}0$.
\end{corollary}
% \begin{remark}
% For Corollary \ref{corollary1}, $d_m$ need \emph{not} be fixed, and it is only required that $d_m=o\left(N^{\frac{\rho^2}{2-\rho^2}}\right)$. In other words, the delay spread can grow with the observation time, as long as it grows ``sufficiently slow" with $N$.
% \end{remark}

% Although Proposition \ref{theorem1} already provides analytical performance guarantees for the MIE, we next present yet another result (in the form of a lower bound), with which we shall obtain a more accurate characterization of the asymptotic error probability in the regime $N\to\infty$.
% \begin{proposition}[Error probability lower bound]\label{theorem2}
% For the same setting and the same asymptotic regime as in Proposition \ref{theorem1}, we have
% \begin{align}
%     \epsilon&\geq \sqrt{\frac{2-\rho^2}{4\pi\rho^2\log(N)}}\cdot\frac{1}{N^{\frac{\rho^2}{2-\rho^2}}}\left(1+o(1)\right)\triangleq \barbelow{\epsilon}(N,\rho)\label{eq:lowerboundmaintheoremdiscretetime},
% \end{align}
% where the $o(1)$ term goes to zero as $N\to\infty$.
% \end{proposition}

% Combining Propositions \ref{theorem1} and \ref{theorem2}, we obtain our main result---the asymptotic error probability rate of decay of the MIE.
\vspace{-0.15cm}Yet another immediate corollary of Theorem \ref{theorem1}, which establishes the asymptotic error probability rate of decay of the MIE, is the following.\vspace{-0.15cm}

\begin{corollary}[Asymptotic error exponent]\label{theorem3}
% For the same setting as in Theorem \ref{theorem1}, we have
\hspace{-0.1cm}In the setting of Theorem \ref{theorem1},
\begin{equation}\label{eq:exactasymptoticerrorexponent}
    \lim_{k\to\infty}-\frac{1}{k}\log_2(\epsilon) = \frac{\rho^2}{2-\rho^2}.
    % \lim_{k\to\infty}-\frac{1}{k}\log_2(\epsilon) = \frac{\rho^2}{2-\rho^2} \; \Longrightarrow \; \epsilon \propto \frac{1}{N^{\frac{\rho^2}{2-\rho^2}}}, \; N\to\infty.
\end{equation}
\end{corollary}
\noindent\textbf{Proof of Corollary \ref{theorem3}} Sandwiching $-\frac{1}{k}\log_2(\epsilon)$ of the left-hand side of \eqref{eq:exactasymptoticerrorexponent} with $-\frac{1}{k}\log_2(\bar{\epsilon}(N,\rho,d_m))$ and $-\frac{1}{k}\log_2\left(\barbelow{\epsilon}(N,\rho)\right)$ from below and above, respectively, and taking $k\to\infty$ gives \eqref{eq:exactasymptoticerrorexponent}. \hfill $\blacksquare$

Observe that even for $\rho=1$ (the ``infinite SNR regime"), $\epsilon$ is \emph{not} zero. Indeed, one of the samples at the ``edges" of the time-interval that are observed by the decoder---but not by the encoder (due to the time-delay uncertainty)---can be greater than the one reported by the encoder.
\vspace{-0.45cm}
\subsection{Proof of Theorem \ref{theorem1}}\label{subsec:proofs}
In order to prove the theorem, we shall use the following lemmas, whose proofs are given below.
\begin{lemma}\label{boundonmaxlowertail}
	Let $\rndj\triangleq \arg \max_{0\leq n\leq N-1}\rndx[n]$, where $\{\rndx[n]\}_{n=0}^{N-1}$ are iid standard normal. Then, for any $\tau\in\reals$,
	\begin{equation}
		\prob{\rndx[\rndj]<\tau} \leq e^{-2^k\left(\frac{\tau}{1+\tau^2}\right)\frac{1}{\sqrt{2\pi}}e^{-\frac{\tau^2}{2}}}.%\triangleq \maxltup(\tau).
	\end{equation}
	Furthermore, if we choose $\tau_*(k)\triangleq \sqrt{2\log(2)k(1-\varepsilon(k))}$, where $\varepsilon(k)\triangleq\frac{1}{\sqrt{k}}=o(1)$, we obtain
	\begin{equation}
		\begin{aligned}
			\prob{\rndx[\rndj]<\tau_*(k)} &\leq e^{-2^{\sqrt{k}}\cdot\frac{1}{\sqrt{2\pi}}\left(\frac{\sqrt{2\log(2)k(1-\varepsilon(k))}}{1+2\log(2)k(1-\varepsilon(k))}\right)}=o\left(2^{-k}\right).\label{superexponentialdecay}
		\end{aligned}
	\end{equation}
\end{lemma}
\begin{lemma}\label{lemmatruncatedgaussian}
	Let $\rndv,\rndz\sim\normal(0,1)$ be independent, and $\rndu\triangleq \min(\rndv,V)$, for some $V\in\reals$. Then, for any $a\in\reals$,
	\begin{equation}\label{eq:exponentialapporxequality}
		\prob{a<\rho\rndu+\bar{\rho}\rndz} \geq \prob{a<\rndv} - Q(V).
	\end{equation}
\end{lemma}
\textit{Proof of Lemma \ref{boundonmaxlowertail}:} For $\tau>0$, we have
\begin{align}
	\prob{\rndx[\rndj]<\tau}&=\prob{\max_{1\leq n \leq N} \rndx[n]<\tau}\\
	&=\prob{\rndx[1]<\tau,\ldots,\rndx[N]<\tau}\\
	&=\prob{\rndx[1]<\tau}^N\\
	&=\left(1-Q(\tau)\right)^N\\
	&\leq\left(1-\left(\frac{\tau}{1+\tau^2}\right)\frac{1}{\sqrt{2\pi}}e^{-\frac{\tau^2}{2}}\right)^N \label{eq:transitiona}\\
	&\leq e^{-N\left(\frac{\tau}{1+\tau^2}\right)\frac{1}{\sqrt{2\pi}}e^{-\frac{\tau^2}{2}}}, \label{eq:transitionc}
\end{align}
where we have used:
\begin{itemize}
	\item $Q(x)\geq \frac{x}{(1+x^2)}\frac{1}{\sqrt{2\pi}}e^{-\frac{x^2}{2}}$ for $x>0$ in \eqref{eq:transitiona} \cite[Eq.~(10)]{gordon1941values}; and
	\item $1-x\leq e^{-x} \; \Rightarrow \; (1-x)^N\leq e^{-xN}$ in \eqref{eq:transitionc}.
\end{itemize}
Choosing $\tau=\tau_*(k)$ gives, after simplifying, \eqref{superexponentialdecay}.
\hfill $\blacksquare$

\noindent\emph{Proof of Lemma \ref{lemmatruncatedgaussian}}: We have,
\begin{align}
	&\prob{a<\rho\rndu+\bar{\rho}\rndz} \\
	&= \Exop\left[\prob{a<\rho\rndu+\bar{\rho}\rndz\mid \rndz}\right]\\
	&=\Exop\left[\prob{\tfrac{a-\bar{\rho}\rndz}{\rho}<\rndu\mid \rndz}\right]\\
	&=\Exop\left[\frac{1}{1-Q\left(V\right)}\int_{\frac{a-\bar{\rho}\rndz}{\rho}}^{V}\frac{1}{\sqrt{2\pi}}e^{-\frac{x^2}{2}}{\rm d}x\right]\\
	&\geq \Exop\left[\int_{\frac{a-\bar{\rho}\rndz}{\rho}}^{\infty}\frac{1}{\sqrt{2\pi}}e^{-\frac{x^2}{2}}{\rm d}x -\int_{V}^{\infty}\frac{1}{\sqrt{2\pi}}e^{-\frac{x^2}{2}}{\rm d}x\right]\label{eq:removetruncatednormalizationfactor}\\
	&=\Exop\left[\prob{\tfrac{a-\bar{\rho}\rndz}{\rho}<\rndv\mid \rndz}\right] - Q(V)\\
	&=\prob{a<\rndv} - Q(V),\label{eq:lasttransitionconvexcombination}
\end{align}
where \eqref{eq:removetruncatednormalizationfactor} is from $[1-Q(V)]^{-1}>1$, and \eqref{eq:lasttransitionconvexcombination} follows from the fact that $\rho^2+\bar{\rho}^2=1$ and that $\rndv$ and $\rndz$ are independent. \hfill $\blacksquare$

\noindent\textit{Proof of Theorem 1:} We start by deriving the upper bound. For brevity in the following derivation, Let $\rndv\sim\normal(0,1)$. Using this notation, we have,
\begin{align}
	&\prob{\left.\widehat{\rndd}\neq \rndd\right|\rndd,\rndx[\rndj]}\\
	&=\prob{\left.\bigcup_{\substack{\ell\in\mathcal{D} 
				\\ \ell\neq \rndd}}\widehat{\rho}(\rndd)< \widehat{\rho}(\ell)\right|\rndd,\rndx[\rndj]}\\
	&\leq 2d_m \prob{\left.\widehat{\rho}(\rndd)< \widehat{\rho}(\ell)\right|\rndd,\rndx[\rndj]} \label{eq:unionbound}\\
	&\leq 2d_m\prob{\rho\rndx[\rndj]<\sqrt{2-\rho^2}\rndv\Bigg|\rndd,\rndx[\rndj]}\label{eq:truncatedbygaussiantrans2}\\
	&=2d_m Q\left(\frac{\rho\rndx[\rndj]}{\sqrt{2-\rho^2}}\right),\label{eq:conditionalqfunction}
\end{align}
where:
\begin{itemize}
	\item In \eqref{eq:unionbound}, we have used the union bound; and
	\item In \eqref{eq:truncatedbygaussiantrans2}, replacing $\rndy[\rndj+\ell]$ by $\rndv$ can only increase the probability. To see this more clearly, we first recall that $\widehat\rho(\ell)$ is merely a scaled version of $\rndy[\rndj+\ell]$. Then, we observe that $\rndy[\rndj+\ell]=\rho\rndx[\rndj-\rndd+\ell]+\bar{\rho}\rndz[\rndj+\ell]$ is a convex combination of a (possibly) one-sided (upper bounded) truncated standard Gaussian RV ($\rndx[\rndj-\rndd+\ell]$) and a standard Gaussian RV ($\rndz[\rndj+\ell]$), which are independent. Since $\rndv$ can be thought of as a convex combination with the same coefficients of two independent standard Gaussian RVs, it is interpreted as replacing the truncated Gaussian $\rndx[\rndj-\rndd+\ell]$ with a standard Gaussian, which can only increase the probability that $\rndy[\rndj+\rndd]<\rndy[\rndj+\ell]$.
\end{itemize}

Now, using the law of total expectation, the conditional upper bound \eqref{eq:conditionalqfunction} and Lemma \ref{boundonmaxlowertail}, we obtain,
\begin{align}
	\prob{\widehat{\rndd}\neq \rndd} &= \Exop\left[\prob{\left.\widehat{\rndd}\neq \rndd\right|\rndd,\rndx[\rndj]}\right] \\
	&\leq \Exop\left[2d_m Q\left(\frac{\rho\rndx[\rndj]}{\sqrt{2-\rho^2}}\right)\right]\\
	&\leq 2d_m Q\left(\frac{\rho\tau_*(k)}{\sqrt{2-\rho^2}}\right) + 2d_m\prob{\rndx[\rndj]<\tau_*(k)}\\
	&=2d_m Q\left(\rho\sqrt{\frac{2\log(N)}{2-\rho^2}}\right)(1+o(1))\\
	&\leq 2d_m \cdot e^{-k\frac{\log(2)\cdot\rho^2}{2-\rho^2}}(1+o(1)),\label{eq:qfunctionbound}
\end{align}
where we recall in particular \eqref{superexponentialdecay}, namely $\prob{\rndx[\rndj]<\tau_*(k)}=o\left(2^{-k}\right)$, and we have used $Q(x)\leq e^{-\frac{x^2}{2}}$ for $x>0$ in \eqref{eq:qfunctionbound}.

For the lower bound, we have,
\begin{align}
	&\prob{\left.\widehat{\rndd}\neq \rndd\right|\rndd,\rndx[\rndj]}\\
	&=\prob{\left.\bigcup_{\substack{\ell\in\mathcal{D} 
				\\ \ell\neq \rndd}}\widehat{\rho}(\rndd)< \widehat{\rho}(\ell)\right|\rndd,\rndx[\rndj]}\\
	&\geq \prob{\left.\widehat{\rho}(\rndd)< \widehat{\rho}(\ell)\right|\rndd,\rndx[\rndj]} \label{eq:onefromunion}\\
	&=\prob{\left.\rndy[\rndj+\rndd]< \rndy[\rndj+\ell]\right|\rndd,\rndx[\rndj]}\\
	&=\Exop\left[\left.\prob{\left.\rndy[\rndj+\rndd]< \rndy[\rndj+\ell]\right|\rndd,\rndx[\rndj],\rndz[\rndj+\rndd]}\right|\rndd,\rndx[\rndj]\right]\\
	% &=\Exop\left[\left.\prob{\left.\rndy[\rndj+\rndd]< \rndy[\rndj+\ell]\right|\rndd,\rndx[\rndj],\rndz[\rndj+\rndd]}\right|\rndd,\rndx[\rndj]\right]\\
	&\geq \Exop\left[\left.\prob{\left.\rndy[\rndj+\rndd]< \rndv\right|\rndd,\rndx[\rndj],\rndz[\rndj+\rndd]}-Q(\rndx[\rndj])\right|\rndd,\rndx[\rndj]\right]\label{eq:usinglemma2}\\
	&=\prob{\left.\rndy[\rndj+\rndd]< \rndv\right|\rndd,\rndx[\rndj]}-Q(\rndx[\rndj])\\
	&=\prob{\left.\rho\rndx[\rndj]< \sqrt{2-\rho^2}\rndv\right|\rndd,\rndx[\rndj]}-Q(\rndx[\rndj])\\
	&=Q\left(\frac{\rho\rndx[\rndj]}{\sqrt{2-\rho^2}}\right)-Q(\rndx[\rndj])\label{eq:lowerboundtouse}
\end{align}
where in \eqref{eq:onefromunion} we have taken only one event of the union of events, and in \eqref{eq:usinglemma2} we have used Lemma \ref{lemmatruncatedgaussian}. Using the law of total expectation, the lower bound \eqref{eq:lowerboundtouse} and Lemma \ref{boundonmaxlowertail}, we obtain,
\begin{align}
	&\prob{\widehat{\rndd}\neq \rndd}\\
	&= \Exop\left[\prob{\left.\widehat{\rndd}\neq \rndd\right|\rndd,\rndx[\rndj]}\right] \\
	&\geq \Exop\left[Q\left(\frac{\rho\rndx[\rndj]}{\sqrt{2-\rho^2}}\right)-Q(\rndx[\rndj])\right]\\
	&\geq \prob{\tau_*(k)<\rndx[\rndj]<\sqrt{2\log(N)}}\\
	&\quad\cdot\Exop\left[\left.Q\left(\frac{\rho\rndx[\rndj]}{\sqrt{2-\rho^2}}\right)-Q(\rndx[\rndj])\right|\tau_*(k)<\rndx[\rndj]<\sqrt{2\log(N)}\right]\\
	&=Q\left(\rho\sqrt{\frac{2\log(N)}{2-\rho^2}}\right)(1+o(1))\\
	&\geq \sqrt{\tfrac{2-\rho^2}{4\pi\rho^2\log(N)}}\cdot e^{-k\frac{\log(2)\rho^2}{2-\rho^2}}(1+o(1)),\label{eq:lowerboundonq}
\end{align}
since we have $\prob{\tau_*(k)<\rndx[\rndj]}=1-o(1)$ and $\prob{\rndx[\rndj]<\sqrt{2\log(N)}}=1-o(1)$, and in \eqref{eq:lowerboundonq} we have used a lower bound on the $Q$-function \cite[Eq.~(13)]{abreu2012very}. \hfill $\blacksquare$

% \noindent\textbf{Proof of Theorem \ref{theorem3}} Sandwiching $-\frac{1}{k}\log_2(\epsilon)$ of the left-hand side of \eqref{eq:exactasymptoticerrorexponent} with $-\frac{1}{k}\log_2(\bar{\epsilon}(N,\rho,d_m))$ and $-\frac{1}{k}\log_2\left(\barbelow{\epsilon}(N,\rho)\right)$ from below and above, respectively, and taking $k\to\infty$ proves the theorem. \hfill $\blacksquare$
\vspace{-0.2cm}
\section{Simulation Results}\label{sec:simulationresults}
\vspace{-0.2cm}
We now present results of a simulation experiment that corroborates our analysis and demonstrates that our method outperforms a rate-distortion (RD) signal compression benchmark and the (possibly naive but) ubiquitous $1$-bit per sample scalar quantization approach (e.g., \cite{weiss2021one,zhang2021direct,ni2023detection}).

\begin{figure}
    \centering
    \includegraphics[width=0.88\columnwidth]{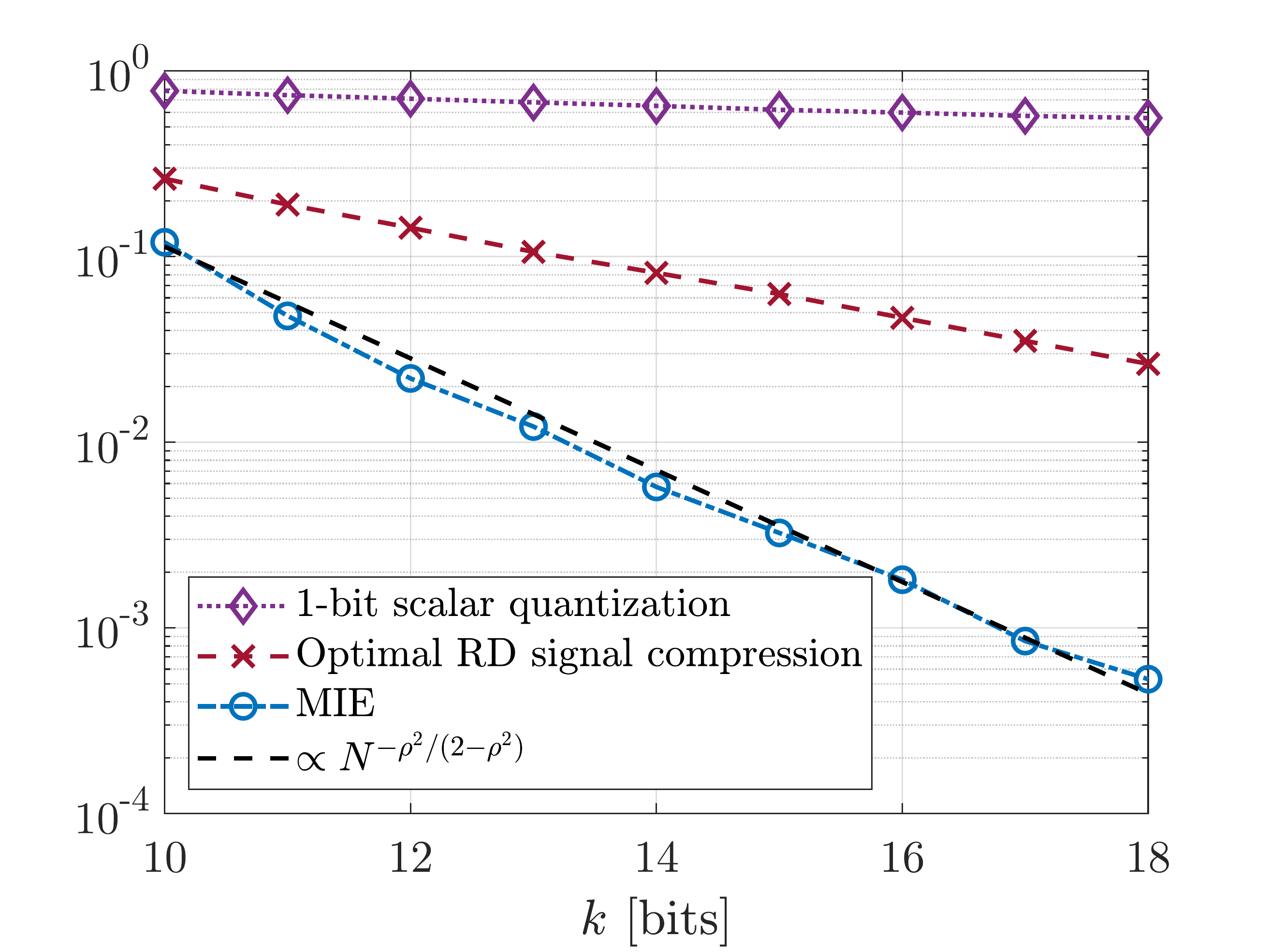}\vspace{-0.3cm}
    % \put(-219,79){\rotatebox{90}{$\mathbb{P}(\epsilon)$}}
    \put(-220,63){\rotatebox{90}{$\mathbb{P}\left(\widehat{\rndd}\neq \rndd\right)$}}
    \caption{Error probability vs.\ the message size (i.e., dimension of $\rndm$) in bits for $\snr=20$dB. The black dashed line is $\widehat{c}\cdot 2^{-k\frac{\rho^2}{(2-\rho^2)}}$, where $\widehat{c}$ is the best least-squares-fitted constant for the empirical curve of the MIE.}
    \label{fig:simulationresults}\vspace{-0.5cm}
\end{figure}

We generate the signal according to model \eqref{eq:model} with $d_{m}=150$ fixed, and compute (i) the MIE \eqref{eq:maxestsimple}; (ii) the cross-correlator \eqref{eq:mledefinition} when $\rndx[n]$ is replaced by $\widehat{\rndx}_{\ratedist}[n]$,\footnote{In this case, this is the still the optimal TDE in the maximum likelihood sense, since $\widehat{\rndx}_{\ratedist}[n]$ is Gaussian \cite[Ch.~10.3.2]{cover1999elements}, and $\widehat{\rndx}_{\ratedist}[n], \rndy[n]$ are jointly Gaussian.} a RD-optimally compressed version thereof, where the distortion measure is the squared error $\left(\rndx[n]-\widehat{\rndx}_{\ratedist}[n]\right)^2$; and (iii) the cross-correlator \eqref{eq:mledefinition} when $\rndx[n]$ is replaced by $\widehat{\rndx}_{\onebit}[n]\triangleq\sign(\rndx[n])$. Figure \ref{fig:simulationresults}, which shows $\prob{\epsilon}$ vs.\ $k$ for $\snr=20$dB, reflects a good empirical fit to our result \eqref{eq:exactasymptoticerrorexponent}, and further demonstrates how our method outperforms any compression scheme that opts to compress a subsequence of $\rndx[n]$ in the MSE sense while ignoring the existence of $\rndy[n]$, and in particular $1$-bit per sample scalar quantization.

\vspace{-0.2cm}
\section{Concluding Remarks}\label{sec:conclusion}
\vspace{-0.2cm}
For the prototype formulation addressed in this work, we present a new joint compression-TDE scheme for distributed systems with communication constraints. Further, we derive its asymptotic error probability (rate of decay) performance. Our scheme has an intuitive interpretation, is simple to implement, and outperforms relevant benchmarks.

While current research efforts are focused on a more refined performance analysis (for this and other regimes of operation) and extensions of the signal model, the proposed approach warrants rethinking the communication efficiency in the context of this specific, but ubiquitous task. Viewing this preliminary work as a seed, we expect that the main concept underlying our approach will be further developed in various directions, and lead to new distributed estimation methods across different domains.

\bibliographystyle{IEEEbib}
\bibliography{./Inputs/refs}

\begin{thebibliography}{10}

\bibitem{viola2003comparison}
F. Viola and W.~F. Walker,
\newblock ``A comparison of the performance of time-delay estimators in medical
  ultrasound,''
\newblock {\em {IEEE} Trans. Ultrason., Ferroelectr., Freq. Control}, vol. 50,
  no. 4, pp. 392--401, 2003.

\bibitem{li2006position}
Y. Li,
\newblock ``Position and time-delay calibration of transducer elements in a
  sparse array for underwater ultrasound imaging,''
\newblock {\em {IEEE} Trans. Ultrason., Ferroelectr., Freq. Control}, vol. 53,
  no. 8, pp. 1458--1467, 2006.

\bibitem{musicki2009mobile}
D. Musicki, R. Kaune, and W. Koch,
\newblock ``Mobile emitter geolocation and tracking using {TDOA} and {FDOA}
  measurements,''
\newblock {\em {IEEE} Trans. Signal Process.}, vol. 58, no. 3, pp. 1863--1874,
  2009.

\bibitem{weiss2022semi}
A. Weiss, T. Arikan, H. Vishnu, G.~B. Deane, A.~C. Singer, and G.~W. Wornell,
\newblock ``A semi-blind method for localization of underwater acoustic
  sources,''
\newblock {\em {IEEE} Trans. Signal Process.}, vol. 70, pp. 3090--3106, 2022.

\bibitem{ziv1969some}
J. Ziv and M. Zakai,
\newblock ``Some lower bounds on signal parameter estimation,''
\newblock {\em {IEEE} Trans. Inf. Theory}, vol. 15, no. 3, pp. 386--391, 1969.

\bibitem{quazi1981overview}
A. Quazi,
\newblock ``An overview on the time delay estimate in active and passive
  systems for target localization,''
\newblock {\em {IEEE} Trans. Acoust., Speech, Signal Process.}, vol. 29, no. 3,
  pp. 527--533, 1981.

\bibitem{ianniello1982time}
J.~P. Ianniello,
\newblock ``Time delay estimation via cross-correlation in the presence of
  large estimation errors,''
\newblock {\em {IEEE} Trans. Acoust., Speech, Signal Process.}, vol. 30, no. 6,
  pp. 998--1003, 1982.

\bibitem{weiss1983fundamental}
A.~J. Weiss and E. Weinstein,
\newblock ``Fundamental limitations in passive time delay estimation--part {I}:
  {N}arrow-band systems,''
\newblock {\em {IEEE} Trans. Acoust., Speech, Signal Process.}, vol. 31, no. 2,
  pp. 472--486, 1983.

\bibitem{weinstein1984fundamental}
E. Weinstein and A. Weiss,
\newblock ``Fundamental limitations in passive time-delay estimation--part
  {II}: {W}ide-band systems,''
\newblock {\em {IEEE} Trans. Acoust., Speech, Signal Process.}, vol. 32, no. 5,
  pp. 1064--1078, 1984.

\bibitem{azaria1984time}
M. Azaria and D. Hertz,
\newblock ``Time delay estimation by generalized cross correlation methods,''
\newblock {\em {IEEE} Trans. Acoust., Speech, Signal Process.}, vol. 32, no. 2,
  pp. 280--285, 1984.

\bibitem{fertner1986comparison}
A. Fertner and A. Sjolund,
\newblock ``Comparison of various time delay estimation methods by computer
  simulation,''
\newblock {\em {IEEE} Trans. Acoust., Speech, Signal Process.}, vol. 34, no. 5,
  pp. 1329--1330, 1986.

\bibitem{carter1987coherence}
G.~C. Carter,
\newblock ``Coherence and time delay estimation,''
\newblock {\em Proc. {IEEE}}, vol. 75, no. 2, pp. 236--255, 1987.

\bibitem{jacovitti1993discrete}
G. Jacovitti and G. Scarano,
\newblock ``Discrete time techniques for time delay estimation,''
\newblock {\em {IEEE} Trans. Signal Process.}, vol. 41, no. 2, pp. 525--533,
  1993.

\bibitem{brandstein1997robust}
M.~S. Brandstein and H.~F. Silverman,
\newblock ``A robust method for speech signal time-delay estimation in
  reverberant rooms,''
\newblock in {\em Proc. Int. Conf. Acoust., Speech, Signal Process.}, 1997,
  vol.~1, pp. 375--378.

\bibitem{bjorklund2003survey}
S. Bj{\"o}rklund,
\newblock {\em A survey and comparison of time-delay estimation methods in
  linear systems},
\newblock M.S. thesis, Dept. Elect. Eng., Linkoping Univ., Linköping, Sweden,
  2003.

\bibitem{benesty2004time}
J. Benesty, J. Chen, and Y. Huang,
\newblock ``Time-delay estimation via linear interpolation and cross
  correlation,''
\newblock {\em {IEEE} Trans. Speech Audio Process.}, vol. 12, no. 5, pp.
  509--519, 2004.

\bibitem{chen2006time}
J. Chen, J. Benesty, and Y. Huang,
\newblock ``Time delay estimation in room acoustic environments: {A}n
  overview,''
\newblock {\em EURASIP J. Appl. Signal Process.}, vol. 2006, pp. 1--19, 2006.

\bibitem{da2014internet}
L. Da~Xu, W. He, and S. Li,
\newblock ``Internet of things in industries: {A} survey,''
\newblock {\em {IEEE} Trans. Ind. Informat.}, vol. 10, no. 4, pp. 2233--2243,
  2014.

\bibitem{liu2020indoor}
M. Liu, L. Cheng, K. Qian, J. Wang, J. Wang, and Y. Liu,
\newblock ``Indoor acoustic localization: {A} survey,''
\newblock {\em Human-centric Computing and Information Sciences}, vol. 10, pp.
  1--24, 2020.

\bibitem{akyildiz2002survey}
I.~F. Akyildiz, W. Su, Y. Sankarasubramaniam, and E. Cayirci,
\newblock ``A survey on sensor networks,''
\newblock {\em {IEEE} Commun. Mag.}, vol. 40, no. 8, pp. 102--114, 2002.

\bibitem{chen2002source}
J.~C. Chen, K. Yao, and R.~E. Hudson,
\newblock ``Source localization and beamforming,''
\newblock {\em {IEEE} Signal Process. Mag.}, vol. 19, no. 2, pp. 30--39, 2002.

\bibitem{wang2003preprocessing}
H. Wang, D. Estrin, and L. Girod,
\newblock ``Preprocessing in a tiered sensor network for habitat monitoring,''
\newblock {\em EURASIP J. Adv. Signal Process}, vol. 2003, no. 4, pp. 1--10,
  2003.

\bibitem{vasudevan2003application}
L. Vasudevan, A. Ortega, and U. Mitra,
\newblock ``Application-specific compression for time delay estimation in
  sensor networks,''
\newblock in {\em Proceedings of the 1st international conference on Embedded
  networked sensor systems}, 2003, pp. 243--254.

\bibitem{fowler2005fisher}
M.~L. Fowler and M. Chen,
\newblock ``Fisher-information-based data compression for estimation using two
  sensors,''
\newblock {\em {IEEE} Trans. Aerosp. Electron. Syst.}, vol. 41, no. 3, pp.
  1131--1137, 2005.

\bibitem{chen2010data}
M. Chen and M.~L. Fowler,
\newblock ``Data compression for multi-parameter estimation for emitter
  location,''
\newblock {\em {IEEE} Trans. Aerosp. Electron. Syst.}, vol. 46, no. 1, pp.
  308--322, 2010.

\bibitem{fuyong2012data}
Q. Fuyong, G. Fucheng, J. Wenli, and M. Xiangwei,
\newblock ``Data compression based on {DFT} for passive location in sensor
  networks,''
\newblock {\em Procedia Engineering}, vol. 29, pp. 3091--3095, 2012.

\bibitem{vargas2018compressed}
E. Vargas, J.~R. Hopgood, K. Brown, and K. Subr,
\newblock ``A compressed encoding scheme for approximate {TDOA} estimation,''
\newblock in {\em Proc. Eur. Signal Process. Conf. (EUSIPCO)}, 2018, pp.
  346--350.

\bibitem{hadar2019distributed}
U. Hadar and O. Shayevitz,
\newblock ``Distributed estimation of {G}aussian correlations,''
\newblock {\em {IEEE} Trans. Inf. Theory}, vol. 65, no. 9, pp. 5323--5338,
  2019.

\bibitem{kochman2021communication}
Y. Kochman and L. Wang,
\newblock ``On the communication exponent of distributed testing for {G}aussian
  correlations,''
\newblock in {\em IEEE Inf. Theory Workshop (ITW)}, 2021, pp. 1--5.

\bibitem{van2004detection}
H.~L. Van~Trees,
\newblock {\em Detection, estimation, and modulation theory, part {I}:
  detection, estimation, and linear modulation theory},
\newblock John Wiley \& Sons, 2004.

\bibitem{gordon1941values}
R.~D. Gordon,
\newblock ``Values of {M}ills' ratio of area to bounding ordinate and of the
  normal probability integral for large values of the argument,''
\newblock {\em Ann. Math. Statist.}, vol. 12, no. 3, pp. 364--366, 1941.

\bibitem{abreu2012very}
G. Abreu,
\newblock ``Very simple tight bounds on the {$Q$}-function,''
\newblock {\em {IEEE} Trans. Commun.}, vol. 60, no. 9, pp. 2415--2420, 2012.

\bibitem{weiss2021one}
A. Weiss and G.~W. Wornell,
\newblock ``One-bit direct position determination of narrowband {G}aussian
  signals,''
\newblock in {\em Proc. IEEE Workshop Stat. Signal Process. (SSP)}, 2021, pp.
  466--470.

\bibitem{zhang2021direct}
G. Zhang, Q. Zhou, T. Zhou, and W. Yi,
\newblock ``Direct position determination with one-bit sampling for
  bandwidth-constrained radar,''
\newblock in {\em IEEE Int. Conf. Signal Image Process. (ICSIP)}, 2021, pp.
  595--600.

\bibitem{ni2023detection}
L. Ni, D. Zhang, Y. Sun, N. Liu, J. Liang, and Q. Wan,
\newblock ``Detection and localization of one-bit signal in multiple
  distributed subarray systems,''
\newblock {\em {IEEE} Trans. Signal Process.}, vol. 71, pp. 2776--2791, 2023.

\bibitem{cover1999elements}
T.~M. Cover and J.~A. Thomas,
\newblock {\em Elements of information theory},
\newblock John Wiley \& Sons, 1999.

\end{thebibliography}

\end{document}